\UseRawInputEncoding

\documentclass[pra,aps,twocolumn,epsfig,nofootinbib,superscriptaddress]{revtex4}
\usepackage{amsmath}
\usepackage{amssymb}
\usepackage[dvips]{graphicx}
\usepackage{dcolumn}
\usepackage{bm}
\usepackage[colorlinks=false,dvipdfm]{hyperref} 
\usepackage{setspace}
\usepackage{amsfonts}
\usepackage{rotating}
\usepackage{makeidx}
\usepackage{amsthm}

\begin{document}

\title{Quantum information masking basing on quantum teleportation}
\author{Wei-Min Shang }
\affiliation{Theoretical Physics Division, Chern Institute of Mathematics, Nankai University, Tianjin 300071, China}

\author{Fu-Lin Zhang }
\email[Corresponding author: ]{flzhang@tju.edu.cn}
\affiliation{Department of Physics, School of science, Tianjin University, Tianjin 300072, China}

\author{Jing-Ling Chen }
\email[Corresponding author: ]{chenjl@nankai.edu.cn}
\affiliation{Theoretical Physics Division, Chern Institute of Mathematics, Nankai University, Tianjin 300071, China}

\date{\today}
\begin{abstract}
The no-masking theorem says that masking quantum information is impossible in a bipartite scenario. However, there exist schemes to mask quantum states in multipartite systems. In this work, we show that, the joint measurement in the teleportation is really a masking process, when the apparatus is regarded as a quantum participant in the whole system.
Based on the view, we present two four-partite maskers and a tripartite masker.
One of the former provides a generalization in arbitrary dimension of the four-qubit scheme given by Li and Wang [Phys. Rev. A 98, 062306 (2018)],
and the latter is precisely their tripartite scheme.
The occupation probabilities and coherence of quantum states are masked in two steps of our schemes. And the information can be extracted naturally in their reverse processes.
\end{abstract}


\maketitle


\section{introduction}
Quantum information not only applies the principles of quantum mechanics to information science, but also provides new perspectives for understanding quantum theory.
Many natural processes in classical information are forbidden by the laws of quantum mechanics, which are termed no-go theorems. For instance, the no-cloning principle  \cite{WW,VS}, no-broadcast principle \cite{HC,AI}, no-deleting principle \cite{AK} and no-superposition principle \cite{MA}.
On the other hand, there are quantum information protocols without classical counterparts, a typical example is quantum teleportation \cite{CG,LZ,DP}.

Recently Modi $et~al$ \cite{KA} found that in a closed system, it is impossible to hide the unknown quantum information into the correlation of bipartite quantum system and observe no original information in the subsystems which is valid for classical information. This is ``no-masking" theorem. This interesting phenomenon has aroused widespread attention. Many meaningful works on this topic have been published \cite{LL,LK,HKC,LLF,LJ,DH}. In  \cite{LW}, authors shows that qudit information masking can be accomplished in 2$d$-qudit or three-qudit system which employed the quantum orthogonal Latin squares \cite{B}. Afterwards, this scenario was generalized to multipartite quantum information masking in multipartite system with the development of $k-$uniform states \cite{SL} which can be obtained from orthogonal arrays \cite{DK}, Latin squares \cite{DD,DZ}, quantum error correcting codes \cite{A} $et~al$.

In this work, we show that,
%
the scheme of teleportation  \cite{CG,LZ}
naturally
contains a process of quantum information masking.
This provides a simple understanding about the process of quantum information masking in multipartite systems.
%
 In the standard teleportation of a qudit \cite{CG}, Alice (the sender) sends an unknown state to Bob (the receiver) by a two-qudit channel without physical transmission of the system itself.
 When Alice measurs the particles in her hands, the original state to be teleported is destroyed.
 Sequentially, she informed Bob of the measurement results utilizing classical communication.
 According to the measurement result, Bob could resume the original qudit state.

Our key point here is that, when Alice's apparatus is regarded as a quantum participant in the whole system,
her measurement provides a four-partite ($d^2 \otimes d \otimes d \otimes d $) scenario to mask the original qubit state.
%
Acturally, Alice's measurement can be divided into two steps:
(1) factorizing the entangled basis by applying a controlled-NOT gate;
(2) measuring the product states.
One can notice that the occupation probabilities of the original state are masked in the first step,
while the coherent terms can be masked by a measurement on principal system.
This leads to our second four-partite masking scenario in a $d \otimes d \otimes d \otimes d $ system,
which returns to the four-qubit scheme given by Li and Wang \cite{LW} when $d=2$.
Furthermore, when $d\geq3$, the coherent terms can be masked by a joint unitary transformation on her two particles,
which can be viewed as a measurement on the  principal system.
%
%
%
Such steps provide an operate process of the tripatite   ($ d \otimes d \otimes d $) masker based on orthogonal Latin squares \cite{LW}.
These results help us understand the absence of a universal masker from a qubit to three qubits,
and intuitively give an unmasking process.

\section{Four-partite masker}

\subsection{$d^2 \otimes d \otimes d \otimes d $ scenario}
Our masking scheme is valid for the unknown states in arbitrary dimension.
Let us start with the quantum teleprotation for qutrit state.
Suppose that the original information contained in the state in qutrit system labeled ``1" given as,
\begin{equation}\label{1}
\begin{aligned}
|\chi_{a}\rangle_{1}=\alpha_{0}|0\rangle+\alpha_{1}|1\rangle+\alpha_{2}|2\rangle,
\end{aligned}
\end{equation}
and quantum channal is the maximum entangled state in two-qutrit system as follow
\begin{equation}\label{2}
\begin{aligned}
|\varphi_{a}\rangle_{23}=\frac{|00\rangle+|11\rangle+|22\rangle}{\sqrt{3}}.
\end{aligned}
\end{equation}
It should be noted that 1,2-partite is in Alice's hand, 3-partite belongs to Bob.
The total state of the three-qutrit system can be written as,
\begin{equation}
\begin{aligned}
|\Psi_{a}\rangle_{123}=\frac{1}{3}\sum_{k,l=0}^{2,2}|\phi_{kl}\rangle_{12} |\chi_{kl}\rangle_{3}.
\end{aligned}
\end{equation}
where
\begin{equation}\label{4}
\begin{aligned}
&|\phi_{kl}\rangle_{12}=\frac{1}{\sqrt{3}}(|l\rangle|0\rangle+\omega^{k}|l\oplus_{n}1\rangle|1\rangle+\omega^{2k}|l\oplus_{n}2\rangle|2\rangle),\\
&|\chi_{kl}\rangle_{3}=\alpha_{l}|0\rangle+\omega^{2k}\alpha_{l\oplus_{3}1}|1\rangle+\omega^{k}\alpha_{l\oplus_{3}2}|2\rangle.
\end{aligned}
\end{equation}
It should be noted that $\omega=e^{i \frac{ 2\pi}{3}}$ is the triple root and we define that $i\oplus_{n}j=(i+j)$mod~$n$.
$\left\vert \psi_{kl}\right\rangle_{12}$  are nine mutually orthogonal two-qutrit maximum entangled state and $|\chi_{kl}\rangle_{3}$ are nine non-orthogonal states.

Up to now, the original information $k$ still contained in 1 system since it is still undisturbed. Moreover, we have no information about the value $k$ from system 2,3 since they are undisturbed too. The partite belong to system 2 shares the maximum entangled state ($\ref{2}$) with the partite belong to system 3, i.e.,  $\rho_{2}=$Tr$_{13}(|\Phi\rangle_{123}\langle\Phi|)=\frac{1}{3}\mathbb{I}$ and  $\rho_{3}=$Tr$_{12}(|\Phi\rangle_{123}\langle\Phi|)=\frac{1}{3}\mathbb{I}$.

If Alice measures her two particles and told Bob the measurement result through classical communication, then Bob could resume the original state by performing the corresponding  unitary operation on his particle thus the original information will be translated to system 3, which known as quantum teleportation. Sequential masking process different from the teleportation. Alice cancels the classical communicating process here and introduces a $9$-level auxiliary system ``0"-system  as a quantum apparatus to measure 1-2 systems and the total four-qutrit state can be wrote as,
\begin{equation}
\begin{aligned}
\left\vert \Psi_{a^{'}}\right\rangle _{0123} =\sum_{k}^{2}|\nu_{kl}\rangle_{0}|\phi_{kl}\rangle_{12}|\chi_{kl}\rangle_{3},
\end{aligned}
\end{equation}
where
\begin{equation}
\begin{aligned}
|\nu_{kl}\rangle_{0}=|3l+k\rangle.
\end{aligned}
\end{equation}
It is distinct that we obtain no information about $k$ from system 0 which is the measurement system. The natural result is
 $\rho_{0}=$Tr$_{123}(|\Phi\rangle_{0123}\langle\Phi|)=\frac{1}{3}\mathbb{I}$.

When Alice introduces the measurement system (0-system), the off-diagonal elements of $\rho_{1}$ vanish and the diagonal elements of $\rho_{1}$ became triplets. Thus we can't observe no information about the original state ($\ref{1}$). By calculating we have $\rho_{1}=$Tr$_{023}(|\Phi\rangle_{0123}\langle\Phi|)=\frac{1}{3}\mathbb{I}$.

Thus we complete the deterministic masking for unknown qutrit states in four-partite system. Similar as the situation for masking unknown qutrit states, we can mask unknown qudit ($d$-level) states into four-partite system by adding $d^{2}$-level measurement system.

Suppose the original information contained in the qudit state given as,
\begin{equation}\label{qunit}
\begin{aligned}
|\chi_{a}\rangle_{1}=\sum_{k=0}^{d-1}\alpha_{k}|k\rangle,
\end{aligned}
\end{equation}
and quantum channal is the maximum entangled state in two-qudit system as follow
\begin{equation}\label{qunit-channel}
\begin{aligned}
|\varphi_{a}\rangle_{23}=\frac{1}{\sqrt{d}}\sum_{l=0}^{d-1}|ll\rangle,
\end{aligned}
\end{equation}
Then we rewrite the total state as,
\begin{equation}\label{total-state}
\begin{aligned}
|\Omega_{a}\rangle_{123}=\sum_{k,l=0}^{d-1,d-1}|\chi_{a}\rangle_{1}|\varphi_{a}\rangle_{23},
\end{aligned}
\end{equation}

In order to mask the original information of value $k$ contained on system 1, Alice cancels the classical communicating process here and introduces a $d^{2}$-level auxiliary system ``0"-system  as a quantum apparatus to measure 1-2 systems and the total four-qudit state can be wrote as
\begin{equation}
\begin{aligned}
\left\vert \Omega_{a^{'}}\right\rangle _{0123} =\frac{1}{d}\sum_{k,l=0}^{d-1,d-1}|\nu_{kl}\rangle_{0}|\phi_{kl}\rangle_{12}|\chi_{kl}\rangle_{3},
\end{aligned}
\end{equation}
where
\begin{equation}
\begin{aligned}
&|\nu_{kl}\rangle_{0}=|dl+k\rangle,\\
&|\phi_{kl}\rangle_{12}=\frac{1}{\sqrt{d}}(\sum_{i=0}^{d-1}\omega^{ik}|f(l)\rangle|i\rangle),\\
&|\chi_{kl}\rangle_{3}=\sum_{i=0}^{d-1}\alpha_{f(l)}\omega^{-ik}|i\rangle,
\end{aligned}
\end{equation}
and
\begin{equation}
\begin{aligned}
&f(l)=l\oplus_{d}i,
\end{aligned}
\end{equation}
 with $\omega=e^{\frac{2\pi i}{d}}$. Thus the original information disappeared in the system 1. By calculating we can derive $\rho_{1}=\frac{1}{d}\mathbb{I}$. Moreover, as the measurement system, system 0 is not disturbed. So it's impossible to obtain any information about the information of value $k$, i.e., $\rho_{0}=\frac{1}{d}\mathbb{I}$ . Since there is no information transmission at present, we can't observe no information from systems 2 and 3. Moreover, the particles in systems 2 and 3 share the maximum entangled state. The natural result is  $\rho_{2}=\rho_{3}=\frac{1}{d}\mathbb{I}$.  Thus we complete the deterministic quantum information masking in four-partite system.

 \subsection{$d \otimes d \otimes d \otimes d $ scenario}
As we know, The larger dimension of the introduced measurement system, the more resource we need to consume. From this viewpoint, we reduce the dimension of the measurement system from $d^{2}$ to $d$ by dividing Alice's measurement into two steps.

Firstly, Alice performs controlled-Not gate $C_{21}$ on $|\Psi_{a}\rangle_{123}$ the total state is
\begin{equation}\label{6}
\begin{aligned}
C_{21}:|\Psi_{a}\rangle_{123}\rightarrow|\Psi_{b}\rangle_{123}=\sum_{k,l=0}^{2,2}|\psi_{kl}\rangle_{12}|\chi_{kl}\rangle_{3}.
\end{aligned}
\end{equation}
Where $C_{21}$ hold the form as the follow
\begin{equation}\label{cnot}
\begin{aligned}
C_{21}=I\otimes|0\rangle_{2}\langle0|+U^{\dagger}\otimes|1\rangle_{2}\langle1|+(U^{\dagger})^{2}\otimes|2\rangle_{2}\langle2|,
\end{aligned}
\end{equation}
with
\begin{equation}
\begin{aligned}
U=|2\rangle\langle0|+|0\rangle\langle1|+|1\rangle\langle2|,
\end{aligned}
\end{equation}
and
\begin{equation}
\begin{aligned}
|\psi_{kl}\rangle_{12}=|l\rangle_{1}(|0\rangle+\omega^{k}|1\rangle+\omega^{2k}|2\rangle)_{2}.
\end{aligned}
\end{equation}

Now Alice cancels the sequential teleportation process.  Since Alice doesn't disturb system 3 which means we have no information about the value of $k$ from system 3,i.e., $\rho_{3}=\frac{1}{3}\mathbb{I}$.

As the control system, system 2 is not disturbed. So we have no information about the value of $k$ from system 2, i.e., $\rho_{2}=\frac{1}{3}\mathbb{I}$.

Let's focus on the controlled system: system 1, which contains the information about the value of $k$. In order to observe the masking process more clearly, we rewrite the total state as,
\begin{equation}\label{8}
\begin{aligned}
|\Psi_{b}\rangle_{123}=&\frac{1}{\sqrt{3}}[\alpha_{0}(|0\rangle_{2}|00\rangle_{31}+|1\rangle_{2}|21\rangle_{31}+|2\rangle_{2}|12\rangle_{31}),\\
                   &+\alpha_{1}(|2\rangle_{2}|10\rangle_{31}+|0\rangle_{2}|01\rangle_{31}+|1\rangle_{2}|22\rangle_{31}),\\
                   &+\alpha_{2}(|1\rangle_{2}|20\rangle_{31}+|2\rangle_{2}|11\rangle_{31}+|0\rangle_{2}|02\rangle_{31})].
\end{aligned}
\end{equation}
By calculating we have $\rho_{1}$ as follow
\begin{equation}\label{11}
\begin{aligned}
\rho_{1}=\frac{1}{3}\left(
  \begin{array}{ccc}
    1 & a &  b \\
    a^{*} & 1 &  c \\
    b^{*} & c^{*} & 1 \\
  \end{array}
\right),
\end{aligned}
\end{equation}
where
\begin{equation}
\begin{aligned}
&a=\alpha_{0}\alpha^{*}_{1}+\alpha_{1}\alpha^{*}_{2}+\alpha_{2}\alpha^{*}_{0},\\
&b=\alpha_{0}\alpha^{*}_{2}+\alpha_{2}\alpha^{*}_{1}+\alpha_{1}\alpha^{*}_{0},\\
&c=\alpha_{0}\alpha^{*}_{1}+\alpha_{1}\alpha^{*}_{2}+\alpha_{2}\alpha^{*}_{0}.
\end{aligned}
\end{equation}
It can be seen that we have no information about the value of $k$ from the diagonal elements of $\rho_{1}$, namely,  the diagonal elements of $\rho_{1}$ are identical. However, some information about the value of $k$ remains in the off-diagonal elements of $\rho_{1}$ since there are nonorthogonal states in system 2-3. In order to further mask the remained information, secondly, Alice introduces an auxiliary $3-$level system labeled ``0" as a quantum apparatus to measure 1-2 partite, then the total state given as
\begin{equation}
\begin{aligned}
|\Psi_{b^{'}}\rangle_{0123}=&\frac{1}{\sqrt{3}}[\alpha_{0}(|0\rangle_{4}|\psi_{0}\rangle_{2}|00\rangle_{31}+|0\rangle_{0}|\psi_{1}\rangle_{2}|21\rangle_{31},\\
                               &+|0\rangle_{0}|\psi_{2}\rangle_{2}|12\rangle_{31}),\\
                    &+\alpha_{1}(|1\rangle_{4}|\psi_{2}\rangle_{2}|10\rangle_{31}+|1\rangle_{4}|\psi_{0}\rangle_{2}|01\rangle_{31},\\
                               &+|1\rangle_{4}|\psi_{1}\rangle_{2}|22\rangle_{31},\\
                    &+\alpha_{2}(|2\rangle_{4}|\psi_{1}\rangle_{2}|20\rangle_{31}+|2\rangle_{4}|\psi_{2}\rangle_{2}|11\rangle_{31},\\
                               &+|2\rangle_{4}|\psi_{0}\rangle_{2}|02\rangle_{31})].
\end{aligned}
\end{equation}
Where
\begin{equation}
\begin{aligned}
|\psi_{k}\rangle_{2}=\frac{1}{\sqrt{3}}(|0\rangle+\omega^{k}|1\rangle+\omega^{2k}|2\rangle).
\end{aligned}
\end{equation}
Thus the off-diagonal elements of $\rho_{1}$ vanish which means we can't observe no information about the value of $k$ from system 1, i.e., $\rho_{1}=\frac{1}{3}\mathbb{I}$. Moreover, as the measurement system, system 0 is undisturbed which means we can't observe no information about the value of $k$ from system 0 too, i.e., $\rho_{0}=\frac{1}{3}\mathbb{I}$.

This scenario shows that we can mask the original information contained in the state on system 1 into four-qutrit system by performing partial process of teleportation (controlled-Not) plus a more economical measurement system.

This scenario can be generated to qudit masking.
The original information contained in qudit state given as ($\ref{qunit}$),
and quantum channel is the maximum entangled state in two-qudit system given as ($\ref{qunit-channel}$).
Alice performs general controlled-Not gate $\mathcal{C}_{21}$ on $|\Omega_{a}\rangle_{123}$ as $(\ref{total-state})$, then the total state is
\begin{equation}\label{general-state}
\begin{aligned}
|\Omega_{b}\rangle_{123}=\frac{1}{\sqrt{d}}\sum_{k,l=0}^{d-1,d-1}\alpha_{k} |k\oplus_{d}1\rangle_{1} |l\rangle_{2}|l\rangle_{3},
\end{aligned}
\end{equation}
where $\mathcal{C}_{21}$ given as
\begin{equation}\label{21}
\begin{aligned}
\mathcal{C}_{21}=\sum_{k,l=0}^{d-1,d-1}|l\rangle_{2}\langle l|\otimes(\mathcal{U}^{\dagger})^{l}.
\end{aligned}
\end{equation}
with the general unitary operator $\mathcal{U}$ given as
\begin{equation}
\begin{aligned}
\mathcal{U}=\sum_{k}|k\oplus_{d}1\rangle\langle k|.
\end{aligned}
\end{equation}

It is distinct that systems 2,3 contain no information about the value $k$, i.e., $\rho_{2}=\rho_{3}=\frac{1}{d}\mathbb{I}$. Moreover the remained information about $k$ is leaked on off-diagonal elements of $\rho_{1}$. In order to further mask the remained information, Alice adds an auxiliary $d-$level system labeled ``0" as a quantum apparatus to measure 1-2 partites and then the total state given as
\begin{equation}
\begin{aligned}
|\Omega_{b^{'}}\rangle_{0123}=\frac{1}{\sqrt{d}}\sum_{k,l=0}^{d-1,d-1}\alpha_{k}|k\rangle_{0}|k\oplus_{d}1\rangle_{1}|l\rangle_{2}|l\rangle_{3},
\end{aligned}
\end{equation}
Thus we can't observe no information about $k$ due to the vanish of the off-diagonal elements of $\rho_{1}$, i.e., $\rho_{1}=\frac{1}{d}\mathbb{I}$. Moreover, as the measurement system, system 0 is not disturbed. So it's impossible to obtain any information about the information of value $k$, i.e., $\rho_{1}=\frac{1}{d}\mathbb{I}$.

Now we analyze the qubit masking in four-qubit system from the unified perspective of teleportation and quantum information masking. Suppose the original information contained in the unknown qubit states as,
\begin{equation}\label{qubit}
\begin{aligned}
|\chi_{b}\rangle_{1}=\alpha|0\rangle+\beta|1\rangle,
\end{aligned}
\end{equation}
and the quantum channel given as
\begin{equation}\label{qubit-channel}
\begin{aligned}
|\varphi_{b}\rangle_{23}=\frac{1}{\sqrt{2}}(|00\rangle+|11\rangle),
\end{aligned}
\end{equation}
Alice performs controlled Not gate $C_{21}$ on the total state $|\Phi_{d}\rangle_{123}=|\chi_{b}\rangle_{1}\otimes|\varphi_{b}\rangle_{23}$ and derive
\begin{equation}
\begin{aligned}
|\Phi_{d^{'}}\rangle_{123}=&\frac{1}{2}[|000\rangle_{012}(\alpha|0\rangle+\beta|1\rangle)_{3},\\
                           &+|001\rangle_{012}(\alpha|0\rangle-\beta|1\rangle)_{3},\\
                           &+|110\rangle_{012}(\alpha|1\rangle+\beta|0\rangle)_{3},\\
                           &-|111\rangle_{012}(\alpha|1\rangle-\beta|1\rangle)_{3}].
\end{aligned}
\end{equation}
In order to catch the masking effect intuitively, we rewrite $|\Phi_{d^{'}}\rangle_{123}$ as
\begin{equation}
\begin{aligned}
&|+\rangle\rightarrow |\Psi_{0}\rangle=\frac{1}{2}(|00\rangle+|11\rangle)\otimes(|++\rangle+|--\rangle),\\
&|-\rangle\rightarrow |\Psi_{1}\rangle=\frac{1}{2}(|00\rangle-|11\rangle)\otimes(|++\rangle-|--\rangle).
\end{aligned}
\end{equation}
where $|+\rangle=\frac{|0\rangle+|1\rangle}{\sqrt{2}}, |-\rangle=\frac{|0\rangle-|1\rangle}{\sqrt{2}}$. It unitary equals to the qubit masking scenario given by Ref. \cite{LW},
\begin{equation}
\begin{aligned}
&|0\rangle\rightarrow |\Psi_{0}\rangle=\frac{1}{2}(|00\rangle+|11\rangle)\otimes(|00\rangle+|11\rangle),\\
&|1\rangle\rightarrow |\Psi_{1}\rangle=\frac{1}{2}(|00\rangle-|11\rangle)\otimes(|00\rangle-|11\rangle).
\end{aligned}
\end{equation}
In other word, our scheme is a generalization of qubit masking in four-qubit scenario given by Ref. \cite{LW}.  Thus we complete the deterministic qudit masking in four-qudit system.

\section{Tripartite masker}

Fewer participants means a more economical scheme. We analyze the quantum information masking in tripartite system from the view of teleportation and found that it's possible to mask the original information by performing another controlled operation instead of introducing the auxiliary system.

 Suppose the original information contained in an unknown qutrit state given as $(\ref{1})$, we employ the quantum channel $(\ref{2})$. Alice performs controlled-Not gate $C_{21}$ as $(\ref{cnot})$ on $|\Psi_{0}\rangle_{123}$,  then we rewrite the total state $(\ref{6})$ as
\begin{equation}
\begin{aligned}
|\Psi_{b}\rangle_{123}=&\frac{1}{\sqrt{3}}[|0\rangle_{1}(\alpha_{0}|00\rangle+\alpha_{1}|22\rangle+\alpha_{2}|11\rangle)_{23},\\
&+|1\rangle_{1}(\alpha_{0}|11\rangle+\alpha_{1}|00\rangle+\alpha_{2}|22\rangle)_{23},\\
&+|2\rangle_{1}(\alpha_{0}|22\rangle+\alpha_{1}|11\rangle+\alpha_{2}|00\rangle)_{23}].
\end{aligned}
\end{equation}
Up to now, we have known that it's impossible obtain no information about the value $k$ from 2,3 systems, i.e., $\rho_{2}=\rho_{3}=\frac{1}{3}\mathbb{I}$. Next, we tend to mask the information remained on the off-diagonal elements of $\rho_{1}$ as $(\ref{11})$. Alice performs another controlled Not gate $C_{12}^{\dagger}$ on $|\Psi_{b}\rangle_{123}$, then the total state hold the form as
\begin{equation}\label{24}
\begin{aligned}
|\Psi_{c}\rangle_{123}=&\frac{1}{\sqrt{3}}[|0\rangle_{1}(\alpha_{0}|00\rangle+\alpha_{1}|22\rangle+\alpha_{2}|11\rangle)_{23},\\
&+|1\rangle_{1}(\alpha_{0}|21\rangle+\alpha_{1}|10\rangle+\alpha_{2}|02\rangle)_{23},\\
&+|2\rangle_{1}(\alpha_{0}|12\rangle+\alpha_{1}|01\rangle+\alpha_{2}|20\rangle)_{23}].
\end{aligned}
\end{equation}
where
\begin{equation}
\begin{aligned}
C_{12}^{\dagger}=|0\rangle_{1}\langle0|\otimes \mathbb{I}+|1\rangle_{1}\langle1|\otimes U^{\dagger}+|2\rangle_{1}\langle2|\otimes U.
\end{aligned}
\end{equation}
Then the off-diagonal elements of $\rho_{1}$ vanish, which means we have no information about the value $k$ in  system 1, i.e.,$\rho_{1}=\frac{1}{3}\mathbb{I}$. Thus we mask the unknown qutrit state into tripartite systems perfectly. Our scheme can be generalized to qudit information masking in tripartite system as follow.

Suppose the original information contained in qudit state given as ($\ref{qunit}$), and quantum channel is the maximum entangled state in two-qudit system given as ($\ref{qunit-channel}$). Similar as the qutrit masking process,  Alice performs a general controlled Not gate $\mathcal{C}_{21}$ given as (\ref{21}). As a natural result, the original qudit information contained on the diagonal elements of $\rho_{1}$ is masked. In other word, the diagonal elements of $\rho_{1}$ are identical. The next target is to mask the information contained on the off-diagonal elements of $\rho_{1}$. Instead of introducing the measurement system, Alice performs another general controlled Not gate $\mathcal{C}_{12}^{\dagger}$ on $|\Omega_{b}\rangle_{123}$. Then the total state given as
\begin{equation}\label{26}
\begin{aligned}
|\Omega_{c}\rangle_{123}=&\frac{1}{\sqrt{d}}\sum_{k,l=0}^{d-1,d-1}|F(k)\rangle_{1}|G(l)\rangle_{2}|l\rangle_{3},
\end{aligned}
\end{equation}
where
\begin{equation}
\begin{aligned}
\mathcal{C}_{12}^{\dagger}=\sum_{k,l=0}^{d-1,d-1}|F(k)\rangle_{1}\langle F(k)|\otimes(\mathcal{W}^{\dagger})^{F(k)},
\end{aligned}
\end{equation}
and
\begin{equation}
\begin{aligned}
\mathcal{W}^{\dagger}=\sum_{l=0}^{n-1}|G(l)\rangle_{2}\langle l|,
\end{aligned}
\end{equation}
with
\begin{equation}
\begin{aligned}
F(k)=k\oplus_{d}1,~~G(l)=d-l~\rm{mod}~d.
\end{aligned}
\end{equation}
Thus we have no original information about the value $k$ due to the vanish of the off-diagonal elements of $\rho_{1}$. By calculating we derive $\rho_{1}=\frac{1}{d}\mathbb{I}$.  Since the original information flow to the correlation instead of  each local system .  We can also obtain no information from 2 system. By calculating we can derive $\rho_{2}=\frac{1}{d}\mathbb{I}$. Moreover, since system 3 is undisturbed, we have no information about value $k$ from system 3, i.e., $\rho_{3}=\frac{1}{d}\mathbb{I}$. Now we complete the deterministic qudit information  masking in three-qudit system. It should be noted that our scheme corresponds to the three-qudit masking scheme proposed by Ref \cite{LW}.

By the way, we can give a intuitive understanding of qubit no-masking in three-qubit system.
Suppose the original information contained in unknown qubit states as ($\ref{qubit}$)
and the quantum channel given as ($\ref{qubit-channel}$).
Alice performs controlled Not gate $C_{21}$ on the total state $|\Phi_{d}\rangle_{123}=|\chi_{c}\rangle_{1}\otimes|\varphi_{c}\rangle_{23}$ and derive
\begin{equation}
\begin{aligned}
|\Phi_{d^{'}}\rangle_{123}=\alpha(|000\rangle+|111\rangle)+\beta(|100\rangle+|011\rangle).
\end{aligned}
\end{equation}
Then Alice implements $C^{\dagger}_{12}$ and obtain
\begin{equation}
\begin{aligned}
|\Phi_{d^{''}}\rangle_{123}=\alpha(|000\rangle+|101\rangle)+\beta(|110\rangle+|011\rangle).
\end{aligned}
\end{equation}
This result equals to the follow process
\begin{equation}\label{33}
\begin{aligned}
&|0\rangle\rightarrow \frac{1}{\sqrt{2}}(|00\rangle+|11\rangle)_{13}|0\rangle_{2},\\
&|1\rangle\rightarrow \frac{1}{\sqrt{2}}(|10\rangle+|01\rangle)_{13}|1\rangle_{2}.
\end{aligned}
\end{equation}
It is distinct that the original information translate from  system 1 to the system 2. We can observe the value $k$ from  system 2 which means a failure masking of qubit information.

We mask the information contained on diagonal elements of $\rho_{1}$ by performing the controlled Not gate $C_{21}$. Successively, we perform unitary operation on system 2 to make the off-diagonal elements of  $\rho_{1}$ disappear, which means we mask the information contained on the off-diagonal elements of $\rho_{1}$. In this way, we realized the deterministic masking in tripartite system. Therefore, for qubit information masking, in addition to the situation ($\ref{33}$), the other situation is
\begin{equation}
\begin{aligned}
A:|\Phi_{d^{'}}\rangle_{123}&\rightarrow|\Phi_{d^{'''}}\rangle_{123},\\
&=\alpha(|000\rangle+|111\rangle)+\beta(|110\rangle+|001\rangle).
\end{aligned}
\end{equation}
This result equals to the follow process
\begin{equation}
\begin{aligned}
&|0\rangle\rightarrow \frac{1}{\sqrt{2}}(\alpha|00\rangle+\beta|11\rangle)_{12}|0\rangle_{3},\\
&|1\rangle\rightarrow \frac{1}{\sqrt{2}}(\beta|00\rangle+\alpha|11\rangle)_{12}|1\rangle_{3}.
\end{aligned}
\end{equation}
It is distinct that the original information is translated from  system 1 to system 3. We can observe the value $k$ from system 3 which means a failure masking of qubit information too.

The above discussion shows that it's impossible to realize the deterministic masking for unknown qubit information in three-qubit system.

The advantage of our scheme is that we can  get insight into each detail of the masking process. Moreover, once we understand the process, we can show the intuitive unmasking process of the original information in each local system by performing the corresponding inverse operations. Next, we show the original information unmasking process.

\section{unmasking}

For quantum information masking in four-partite system, we can obtain a state $|\psi_{kl}\rangle$ in 1-2 system  when measured on system 0. Based on the measurement result, we unmasks the original information to system 1 by performing unitary operation $A$ on $|\psi_{kl}\rangle$ where
\begin{equation}
\begin{aligned}
A=\mathcal{U}^{l}\otimes (V^{\dag})^{k},
\end{aligned}
\end{equation}
with
\begin{equation}
\begin{aligned}
V=\sum_{l}\omega^{l}|l\rangle\langle l|.
\end{aligned}
\end{equation}
On this foundation, Alice implements the standard teleportation, and transmits the original information to the system 3. This means we unmask the original information to the system 3.

For qudit information masking in tripartite system, we mask the original information in $|\Omega_{c}\rangle_{123}$ of $(\ref{26})$. As the analogical situation of the qutrit unmasking process, we can apply unitary operation ($\mathcal{C}^{\dagger}_{21}\mathcal{C}_{12}$) on $|\Omega_{c}\rangle_{123}$ to unmask the original information to system 1. Similarly, we can unmasks the original information to the system 2 by performing the unitary operation ($\mathcal{C}^{\dagger}_{32}\mathcal{C}_{23}$) or to the system 3 by performing the unitary operation ($\mathcal{C}^{\dagger}_{31}\mathcal{C}_{13}$) on $|\Omega_{c}\rangle_{123}$ due to the rotating symmetry. Thus we complete the qudit information unmasking process.
\section{summary}
Teleportation is a quantum information process without classical counterparts, which serves as an important example for the most intriguing uses of entanglement.
Such processes can be veiwed as masking and extracting quantum information, when the apparatus of the sender is regarded as a part the whole system.
We convert the teleportation into two four-partite and a tripartite scenarios for quantum information masking.
One of the four-partite masker provides a generalization in arbitrary dimension of the four-qubit scheme given by Li and Wang \cite{LW}, and the tripartite one is precisely their tripartite scheme.
Our operate process show that the occupation probabilities and coherence of quantum states are masked in two measurements.
The absence of a universal masker from a qubit to three qubits can be understood intuitively.
\begin{acknowledgments}
 This work was supported by Nankai Zhide Foundation,
 Tianjin Research Innovation Project for Postgraduate Students (Grant No. 2020YJSB155)
 and   National Natural Science Foundations of China (Grant Nos. \ 11675119, \ 11875167 and \ 12075001)
\end{acknowledgments}
\vspace{8mm}


\end{document}